\documentclass[dvipdfmx]{ptephy_v1}%%%%%% to generate preprint number with ptep logo

\preprintnumber{XXXX-XXXX} %%% %%% Insert preprint number here

\usepackage{graphicx}
\usepackage{color}

\newcommand{\eqsref}[1]{eq.(\ref{#1})}
\newcommand{\figref}[1]{Fig.\ref{#1}}
\newcommand{\secref}[1]{section \ref{#1}}
\newcommand{\tabref}[1]{Table \ref{#1}}

\begin{document}

\title{Revisiting the oscillations in the CMB angular power spectra at $\ell\sim120$ in the Planck2015 data}

%%%% To generate auto affiliation numbers please use \author{}\affil{} command

\author{Koichiro Horiguchi}
\affil{Department of Physics and Astrophysics, Nagoya University, Nagoya-city, Aichi 464-8602, Japan \email{horiguchi.kouichirou@h.mbox.nagoya-u.ac.jp}}

\author{Kiyotomo Ichiki}
\affil{Kobayashi-Maskawa Institute for the Origin of Particle and the Universe, Nagoya University, Nagoya-city, Aichi 464-8602, Japan}

\author{Jun'ichi Yokoyama}
%\author[3]{Insert fourth author name here} %%% Use optional bracket [3] to change the respective address
\affil{Research Center for the Early Universe (RESCEU), Graduate School
 of Science, The University of Tokyo,
 Tokyo 113-0033, Japan}
 \affil{
Department of Physics, Graduate School of Science, The University of Tokyo, Tokyo, 113-0033, Japan}
\affil{Kavli Institute for the Physics and Mathematics 
of the Universe (Kavli IPMU),
UTIAS, WPI, The University of Tokyo, Kashiwa, Chiba, 277-8568, Japan}

%\author{Insert last author name here\thanks{These authors contributed equally to this work}}
%\affil{Insert last author address here}

%%% To include the collaborator name... Please use the command "\collaborator"
%%% For example: \collaborator{ATLAS Collaboration}

\begin{abstract}%
While the observed
nearly scale-invariant initial power spectrum is regarded as
one of the favorable evidence of the standard inflationary cosmology,
precision observations of 
 the Cosmic Microwave Background (CMB) anisotropies also suggest 
possible existence of nontrivial features such as those observed
around  multipoles $\ell\sim120$ by WMAP.  Here, we  examine the Planck 
data and investigate the effects of these features on the cosmological 
parameter estimation performing the Markov-Chain Monte-Carlo (MCMC) 
analysis. We find that the features exist in the Planck data at the 
same position as the case of the WMAP data but they do not affect 
the cosmological parameter estimation significantly.
\end{abstract}

\subjectindex{xxxx, xxx}

\maketitle

%----------Introduction----------%
\section{Introduction} 
In the modern cosmology
the origin of the large-scale structure of the universe is 
attributed to
 tiny initial quantum fluctuations 
 \cite{Mukhanov:1981xt,1982PhLB..115..295H, 1982PhRvL..49.1110G,1982PhLB..117..175S} 
produced during the inflation epoch
 \cite{STAROBINSKY198099,PhysRevD.23.347,1981MNRAS.195..467S} (for a
 review of inflation, see, {\it e.g.} \cite{Sato:2015dga}).
 The simplest models of  inflation predict nearly scale-invariant 
initial power spectrum of curvature perturbations
which can be characterized by the amplitude and the power-law spectral
index \cite{1982PhLB..115..295H, 1982PhLB..117..175S,
1982PhRvL..49.1110G}. 
Motivated by this theoretical background as well as for the sake of
simplicity, the power-law initial spectrum has been tested by precision
observations of comic microwave background (CMB) radiation by
 WMAP \cite{2009ApJS..180..330K} and
Planck \cite{2016A&A...594A..13P}, and shown to provide a good fit.
Furthermore values of the cosmological parameters have been determined
by CMB data mostly under the assumption of the power-law initial spectrum.
From purely observational point of view, 
however, the shape of the initial spectrum of our Universe should be
determined from observational data alone, without any theoretical
prejudice.

In fact, much work has been done to reconstruct the primordial spectrum
from observed CMB data
using a number of methods such as Markov Chain Monte Carlo (MCMC)
analysis of parameterized spectrum \cite{2010PhRvD..81h3010I,2013JCAP...12..035H,2016JCAP...08..028R,2016arXiv161203490G}, cosmic inversion \cite{Matsumiya:2001xj,Matsumiya:2002tx,Kogo:2003yb,Kogo:2005qi,2009PhRvD..79d3010N}, and maximum
likelihood reconstruction methods \cite{2006MNRAS.367.1095T,2009PhRvD..79d3010N,2014A&A...571A..22P}, to name a few.
As a result, a number of possible features imprinted on otherwise
power-law spectrum have been reported in the literatures.

Although it is difficult to interpret them in the context of
inflationary cosmology, namely, to judge if models predicting featured
spectra are really necessary, presence of spectral features can affect
the estimation of the cosmological parameters of the homogeneous and
isotropic background universe.  Then unless we fully incorporate such
features in the parameter analysis we may not obtain correct values of
cosmological parameters.

The purpose of this paper is to examine the feature found around the
multipole $\ell \sim 120$ \cite{2010PhRvD..81h3010I} based on TT and TE
data of  5 year WMAP observation (WMAP5) \cite{2009ApJS..180..330K} is persistent in
the latest Planck data \cite{2016A&A...594A..13P}, which now includes more polarization
data as well,
and how its presence affects the estimation of other cosmological
parameters.  Here T refers to temperature anisotropy and E to the E-mode
polarization. 

Note that this feature has been observed 
 in all sky regions in the WMAP data \cite{2013PhRvD..87b3008K}.
%
%the validity of the power-law initial power spectrum should 
% be tested by observational data \cite{2008PhRvD..78l3002N,2009PhRvD..79d3010N,2012JCAP...06..006V,2013JCAP...07..031H,2013JCAP...12..035H,2014JCAP...01..025H,2014JCAP...11..011H}. The previous 
% studies have looked into the TT and TE correlations of the 
%WMAP5 data searching 
% irregular features by using the two non-parametric methods, namely cosmic inversion and 
% likelihood based methods  
% \cite{2008PhRvD..78l3002N,2009PhRvD..79d3010N}. Both of the two have found features 
% at multipoles $\ell\sim120$ and these features have been analyzed in \cite{2010PhRvD..81h3010I} 
% using several fitting models. Though they could not  be explained in the context of 
% the standard slow-roll inflation \cite{2011JCAP...12..008K}, 
If it is a real cosmological feature, it 
should be found in the Planck data as well and 
 we should take these features into account in cosmological analysis.
 %\Red{
 %Add 1.[examples of fine structures at other scales] and 2.[the mechanism of the structure production] around here.

 In addition, previous works
 \cite{2013JCAP...12..035H,2014JCAP...01..025H,2014JCAP...11..011H} have
 reconstructed some local features 
from recent observational data on other scales.
 These features probably originate from fine distortions of the initial power spectrum which are induced in the inflation epoch. There are a number of inflation mechanisms that could induce such distortions, for instance, the dynamics of a heavy field in multi-field inflation \cite{2011JCAP...01..030A,2012JCAP...05..008C,2012JCAP...10..040G,2012JCAP...11..036S,2015JHEP...08..115G}, brane wrapped inflation \cite{2013JCAP...02..005K}, features from the modified slow-roll inflation potential called local feature \cite{2016EPJC...76..385C,2016arXiv161203490G}, change in the sound velocity in the inflation epoch \cite{2011PThPh.125.1035N,2014PhRvD..90b3511A,2016arXiv161110350T}, and the non-local inflationary feature from wiggly whipped inflation \cite{2016JCAP...09..009H}. Analyses of these features could be a good index for inflation models.%}
 
%\Red{mention about the features in the other 
%scale e.g. ell<50 or large range of ell}
 
% In this paper, we have searched for the features 
%at $\ell\sim120$ in the latest Planck data.
% The aim is to know whether or not they exist in the Planck data 
% and affect the estimation of the cosmological parameters in the MCMC analysis. 
 %\Red{%Add 3.[the concrete description of effects of features on the parameter analysis in WMAP epoch citing \cite{2010PhRvD..81h3010I}, e.g. about affected parameters $(A, n_s)$ and how they are affected.]

 The previous study \cite{2010PhRvD..81h3010I} estimated the
 cosmological 
parameters with and without the features in the MCMC analysis, and they found 
that these features affect particularly the estimation of the parameters 
for the initial power spectrum, the amplitude $A$ and the spectral 
index $n_s$, using the 5yr WMAP data. In their analysis, the resultant 
amplitude $A$ turned out to be smaller and the spectral index $n_s$
 larger than the ones based on the power-law initial power spectrum 
without features. We need to examine that these trends could be seen or not, using the Planck data. %}
 We pay particular attention to the impact of the newly released E-mode auto-correlation data 
 on analyzing the features.

The rest of the paper is organized as follows. 
In \secref{sec:method}, we will explain the parameterization of the
features and the setup of the analysis. To understand the effects of the
features, here we will perform an MCMC analysis using the Planck
un-binned angular power 
spectrum data under the standard $\Lambda$CDM cosmology. In
\secref{sec:result}, we present the results of the analysis and discuss
the existence of the features and their impacts on the parameter
estimation. Then we check the effects of the Planck polarization data on
the analysis of the features. 
Section 4 is devoted to conclusion.

%----------Oscillation model----------%
\section{Method}
\label{sec:method}
We examine whether the features around multipole $\ell \sim 120$ 
found in the WMAP data  \cite{2010PhRvD..81h3010I,
2008PhRvD..78l3002N,2009PhRvD..79d3010N, 2011JCAP...12..008K,
2013PhRvD..87b3008K} are persistent in the Planck data in terms of
 MCMC analysis using the COSMOMC \cite{2002PhRvD..66j3511L} code.
In this section, we give our parameterized model of the features and explain the setup of the analysis.

\subsection{Model of the features}
In the standard cosmology, we usually
 adopt the power law initial power spectrum,
\begin{equation}
\label{eq:ips}
P(k)=\frac{k^3P_\zeta(k)}{2\pi^2}=A\left(\frac{k}{k_0}\right)^{n_s-1},
\end{equation} 
where $P_\zeta(k)$ is the power spectrum of primordial curvature
fluctuation
in the comoving gauge, 
$A$ is the amplitude of the fluctuation, $n_s$ is the spectral index and $k_0=0.05{\rm Mpc}^{-1}$ is the pivot scale. 
Here we assume that the features in the CMB power spectra are originated in the initial power spectrum and explain the features modifying the initial power spectrum \eqsref{eq:ips}.
Besides this overall power-law component, we incorporate a feature
around a comoving wavenumber $k_\ast$ following the previous work
 \cite{2010PhRvD..81h3010I}, where several models of features in the
 primordial spectrum were tested.
Among them we adopt the following functional form
\begin{eqnarray}
\label{eq:ips_wo}
P(k)=A\left(\frac{k}{k_0}\right)^{n_s-1}+B\left(\frac{k}{k_0}\right)^{n_s-1}{\rm exp}\left(-\frac{(k-k_*)^2}{\kappa^2}\right){\rm cos}\left(\frac{k-k_*}{\kappa}\right),
\end{eqnarray}
which reproduced the WMAP5 data the best.
Here $B$, $\kappa$ and $k_*$ correspond to the amplitude, the width and the position of the oscillations, respectively. The product of $k_*$ and the angular diameter distance to the last scattering surface $d_{\rm ang}$ is equivalent to the position of the features in the multipole space 
($k_*d_{\rm ang}\sim \ell$). We call the parameters $B$, $\kappa$ and $k_*d_{\rm ang}$ the feature parameters hereafter.
%\TODO ?
We will use these two initial power spectra \eqsref{eq:ips} and \eqsref{eq:ips_wo} to investigate the statistical significance of the features below.
%----------Analysis setuo----------%
\subsection{Analysis setup}
We perform the MCMC analysis under the standard $\Lambda$CDM cosmology. Here we run the feature parameters in addition to the cosmological parameters with flat priors. We set the range of the feature parameters as $0\leq10^{10}B\leq150$,  
$1\leq10^4\kappa d_{\rm ang} \leq30$ and $100\leq k_*d_{\rm ang}\leq140$, for the amplitude, the width and the position, respectively. 
To understand the effects of the polarization data on the estimation of
the cosmological parameters and the feature parameters, we performed the
MCMC analysis for two different data sets. 
One consists only of the TT auto-correlation data, and the other 
 of the combined data which contains the TT, EE auto-correlation and the TE cross-correlation data of the un-binned Planck2015 angular power spectra.

%----------Result and Discussion----------%
\section{Result \& Discussion}
\label{sec:result}
%Here we have performed MCMC analysis for model1 and model2 using Planck 2015 TT and TT, TE, EE combined data. 
In this section, we  show the result of the analysis. We have performed
several MCMC analyses to compare the standard initial power spectrum
\eqsref{eq:ips} and the initial power spectrum with oscillations given
by \eqsref{eq:ips_wo}. 
%We have analyzed the only TT auto-correlation data and the TT, TE 
%and EE combined data to see the effects of the polarization data 
%on the estimation of the parameters. 
Here we show the  best fit values in \tabref{table:cosmo_params_best} and the mean values in  \tabref{table:cosmo_params_mean} for the cosmological and the feature parameters, and $\chi^2$ values in \tabref{table:chi_square}.

\begin{table}[!h]
\caption{Best fit cosmological and feature parameters for Planck 2015 TT
 data and for TT, TE, EE combined data. Here $\tau$ is the optical depth, $H_0=100h\ [{\rm km/s/Mpc}]$ is the Hubble parameter, $\Omega_bh^2$ and $\Omega_ch^2$ represent the density parameters of baryon and cold dark matter, respectively. We investigated models with the standard initial power spectrum given by \eqsref{eq:ips}, the oscillating initial power spectrum by \eqsref{eq:ips_wo} and the fixed oscillation models in which we used the oscillating initial power spectrum by \eqsref{eq:ips_wo} with the feature parameters being fixed to the best fit values.}%%%Table caption goes here
\label{table:cosmo_params_best}
\centering
\begin{tabular}{|c||c|c||c|c|c|}%%%The number of columns has to be defined here
\hline
  & Standard & with oscillations & Standard & with oscillations & fixed oscillations\\ %%%% Table body
 & TT & TT & TTTEEE & TTTEEE & TTTEEE\\
\hline
$10^{10}A$ &$21.6$ & $21.6$& $21.2$ & $21.3$ & $21.4$\\%%%% Table body
\hline
$n_s$ &$0.969$ &$0.970$ & $0.965$ & $0.967$ & $0.967$\\%%%% Table body
\hline
$\tau$ &$0.0719$ & $0.0707$& $0.0613$&$0.0644$ & $0.0649$\\%%%% Table body
\hline
$H_0$ &$68.0$ &$68.0$ & $67.4$& $67.6$ & $67.6$\\%%%% Table body
\hline
$\Omega_bh^2$ &$0.0223$ &$0.0223$ & $0.0222$ & $0.0223$ &$0.0223$\\%%%% Table body
\hline
$\Omega_ch^2$ &$0.118$& $0.118$& $0.119$ & $0.119$ &$0.119$\\%%%% Table body
\hline
$10^{10}B$ &- &$46.3$ & -&$37.4$ &-\\%%%% Table body
\hline
$10^4\kappa$ &-&$3.09$ & -&$3.14$ &-\\%%%% Table body
\hline
$k_*d_{\rm ang}$ &-& $124.0$& -&$123.5$ &-\\%%%% Table body
\hline
\end{tabular}
\end{table}%%%End of the table

\begin{table}[!h]
\caption{
The same as \tabref{table:cosmo_params_best}, but here we denote the
 mean cosmological and feature parameters for Planck 2015 TT data and
 for TT, TE, EE combined data with $1\sigma$ errors.}%%%Table caption goes here
\label{table:cosmo_params_mean}
\centering
\begin{tabular}{|c||c|c||c|c|c|}%%%The number of columns has to be defined here
\hline
  & Standard & with oscillations & Standard & with oscillations & fixed oscillations\\ %%%% Table body
 & TT & TT & TTTEEE & TTTEEE & TTTEEE\\
\hline
$10^{10}A$ &$21.5^{+0.9}_{-1.0}$ & $21.6^{+0.9}_{-1.1}$& $21.2^{+0.7}_{-0.6}$ & $21.3^{+0.6}_{-0.8}$ & $21.3^{+0.6}_{-0.8}$\\%%%% Table body
\hline
$n_s$ &$0.968^{+0.007}_{-0.008}$ &$0.970^{+0.008}_{-0.008}$ & $0.965^{+0.005}_{-0.005}$ & $0.966^{+0.005}_{-0.005}$ & $0.966^{+0.005}_{-0.005}$\\%%%% Table body
\hline
$\tau$ &$0.0694^{+0.0229}_{-0.0261}$ & $0.0721^{+0.0246}_{-0.0267}$& $0.0609^{+0.00171}_{-0.00173}$&$0.0634^{+0.0170}_{-0.0184}$ & $0.0631^{+0.0171}_{-0.0192}$\\%%%% Table body
\hline
$H_0$ &$68.0^{+1.1}_{-1.3}$ &$68.1^{+1.2}_{-1.2}$ & $67.4^{+0.7}_{-0.7}$& $67.5^{+0.7}_{-0.7}$ & $67.5^{+0.7}_{-0.7}$\\%%%% Table body
\hline
$\Omega_bh^2$ &$0.0223^{+0.0002}_{-0.0003}$ &$0.0223^{+0.0003}_{-0.0003}$ & $0.0222^{+0.0002}_{-0.0001}$ & $0.0222^{+0.0002}_{-0.0002}$ &$0.0222^{+0.0002}_{-0.0002}$\\%%%% Table body
\hline
$\Omega_ch^2$ &$0.118^{+0.003}_{-0.002}$& $0.118^{+0.003}_{-0.003}$& $0.119^{+0.002}_{-0.001}$ & $0.119^{+0.002}_{-0.002}$ &$0.119^{+0.002}_{-0.001}$\\%%%% Table body
\hline
$10^{10}B$ &- &$43.2^{+11.6}_{-43.2}$ & -&$32.5^{+13.8}_{-11.5}$ &-\\%%%% Table body
\hline
$10^4\kappa$ &-&$4.32^{+0.18}_{-2.57}$ & -&$3.57^{+0.51}_{-1.04}$ &-\\%%%% Table body
\hline
$k_*d_{\rm ang}$ &-& $123.1^{+2.7}_{-0.9}$& -&$123.0^{+0.9}_{-0.1}$ &-\\%%%% Table body
\hline
\end{tabular}
\end{table}%%%End of the table

%----------table----------%
\begin{table}[!h]
\caption{Best fit $\chi^2$ for the same models as \tabref{table:cosmo_params_best}.}%%%Table caption goes here
\label{table:chi_square}
\centering
\begin{tabular}{|c||c|c||c|c|c|}%%%The number of columns has to be defined here
\hline
Model& Standard & with oscillations & Standard & with oscillations & fixed oscillations\\ %%%% Table body
data& TT & TT & TTTEEE & TTTEEE & TTTEEE\\
\hline
$\chi^2$ &8431.9 & 8418.2& 24190.9 & 24174.8 &24174.8\\%%%% Table body
\hline
$\Delta\chi^2$ &- & -13.7& - & -16.1 & -16.1\\%%%% Table body
\hline
\end{tabular}
\end{table}%%%End of the table

\subsection{Features in the Planck data}
\begin{figure}[!h]
\centering
\includegraphics[width=5in]{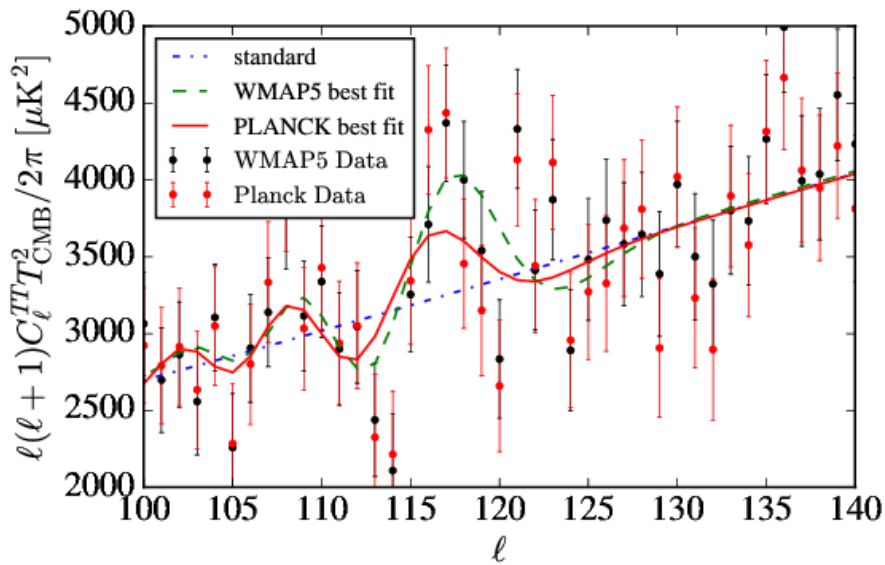}
%%%call your figure name in the place "figurename.eps"
\caption{The angular power spectrum of the TT auto-correlation. The black (red) dots and bars show the  WMAP5 (Planck2015) data points and their error bars. The green dashed line, the blue dot dashed line and the red solid line represent the best fitting model for the TT and TE combined data of WMAP5 using the oscillating initial power spectrum given by \eqsref{eq:ips_wo}, for the TT, TE and EE combined data of Planck2015 using the standard initial power spectrum by \eqsref{eq:ips} and the oscillating initial power spectrum by \eqsref{eq:ips_wo}, respectively. }
\label{fig:ClTTwmap5}
\end{figure}

\begin{figure}[!h]
\centering
\includegraphics[width=5in]{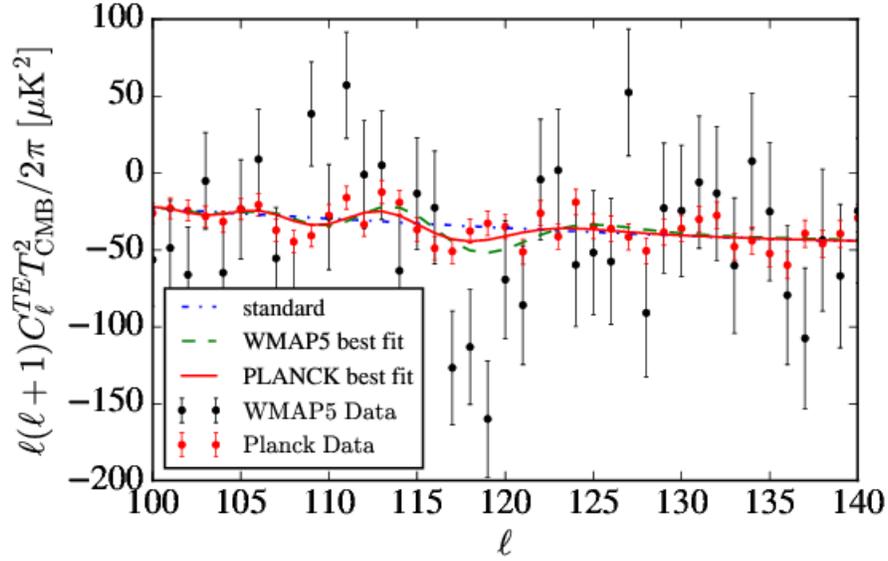}
%%%call your figure name in the place "figurename.eps"
\caption{ Same as \figref{fig:ClTTwmap5} but for the Planck TE cross-correlation data.}
\label{fig:ClTEwmap5}
\end{figure}

Let us focus on the feature parameters first. The previous work 
\cite{2010PhRvD..81h3010I} 
has shown that the best fit values are 
$10^{10}B=55.6$, $10^{4}\kappa=3.58$ and $k_*d_{\rm ang}=124.5$ using 
the TT auto-correlation and TE cross-correlation spectra of the 5yr WMAP 
 data. Comparing the new results shown in 
 \tabref{table:cosmo_params_best} and
 \tabref{table:cosmo_params_mean} with these values, 
we may say that the new result is basically in good agreement with the
previous one \cite{2010PhRvD..81h3010I}.
The main difference in the feature parameters between the WMAP5 data and
the Planck data is the amplitude of the oscillatory feature. 
To understand this difference, let us compare the data points of the 
angular power spectra of WMAP5 and Planck2015. Figure
\ref{fig:ClTTwmap5} shows that at first glance
the data points of the TT auto-correlation are almost the same between
the WMAP5 and the Planck2015 data,
as the both measurements are cosmic-variance-limited in these multipoles.
Therefore these points are not responsible for the change of the feature 
parameters, although it is interesting to note that Planck2015 and WMAP5 data
are deviated from each other well beyond their measurement errors.
%in the TE cross-correlation data.
%\underline{(check required!)}. 
In \figref{fig:ClTEwmap5}, where we show the TE cross-correlation data, 
we can see that both the amplitude of the oscillations and errors have
 become smaller in the Planck data. Therefore we conclude that this
 improved
 TE cross-correlation data makes the value of the amplitude parameter
 $B$ somewhat
smaller for the Planck2015 data compared with the WMAP5 data. 

%----------ClTT, TE, EE----------%

\begin{figure}[!h]
\centering
\includegraphics[width=5in]{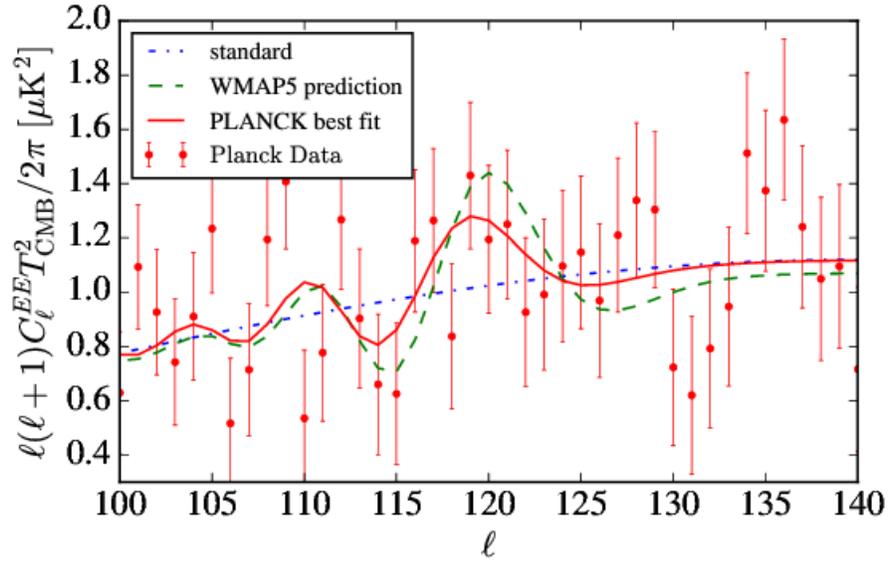}
%%%call your figure name in the place "figurename.eps"
\caption{ Same as \figref{fig:ClTTwmap5} but for the Planck EE auto-correlation data.}
\label{fig:ClEE}
\end{figure}

\begin{figure}[!h]
\centering
\includegraphics[width=5in]{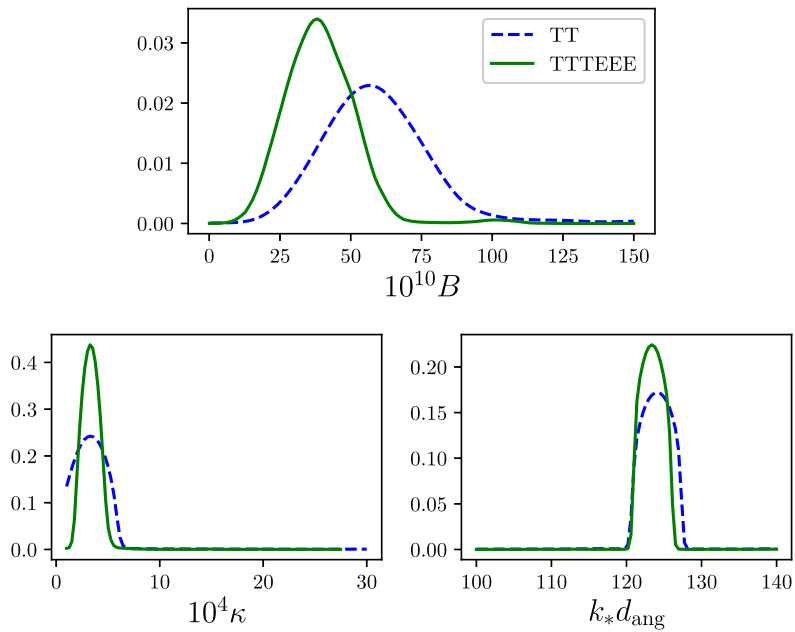}
%%%call your figure name in the place "figurename.eps"
\caption{The normalized likelihoods of the amplitude of the oscillations (top), the width (bottom left) and the position (bottom right) from the TT auto-correlation data of Planck2015 (blue dashed line) alone and  the TT, TE and EE combined data of Planck2015 (green solid line).}
\label{fig:triplotpars}
\end{figure}

The E-mode polarization auto-correlation data have been newly released 
by the Planck collaboration in 2015. In \figref{fig:ClEE}, we can see
the 
prediction by the WMAP5 data fits pretty well to the Planck EE power 
spectrum. For  more accurate understanding of features in the 
polarization data, we show  the likelihoods of the feature 
parameters in \figref{fig:triplotpars}. There we 
show a comparison between the likelihood of the feature parameters 
only from the TT auto-correlation data and the one from the TT, TE 
and EE combined data with smaller uncertainties. 
We can see that all of the likelihoods are 
sharpened, and the amplitude parameter becomes slightly smaller if 
we use the combined data. These indicate that there are features 
with a smaller amplitude, but with the same width and at the same 
position in the polarization data as in the temperature data.

\tabref{table:chi_square} shows the $\chi^2$ values are improved 
by using the oscillating initial power spectrum, and the improvement 
is more than twice the  number of the added parameters. In  view
of Akaike's Information Criterion \cite{akaike1974}, 
this implies that the oscillating initial power spectrum
(\ref{eq:ips_wo}) 
is the better model to describe the primordial spectrum of curvature
perturbation of our Universe.
This does not necessarily mean that inflation model which produced our
Universe as it is today must realize such a featured spectrum as a mean
value because theoretical cost to realize such an ad hoc spectrum cannot
be taken into account in Akaike approach.  What is important here is the
fact that our Universe is imprinted with such a feature whether it was
generated as a result of a non-standard inflation model or as a very
rare realization of conventional inflation model predicting a simple
power-law spectrum.  In either case, in order to determine precise values of
cosmological parameters of our Universe, we should incorporate these
features in the MCMC analysis.

If they are real imprinted features, they can be seeds of 
the large scale structure of the universe, 
and we will be able to see them in the galaxy correlations \cite{2016JCAP...10..041B}.

%-----------------likelihoods
%-----------------TT & TTTEEE

\subsection{The effects on the cosmological parameters}
\begin{figure}[!h]
\centering
\includegraphics[width=5in]{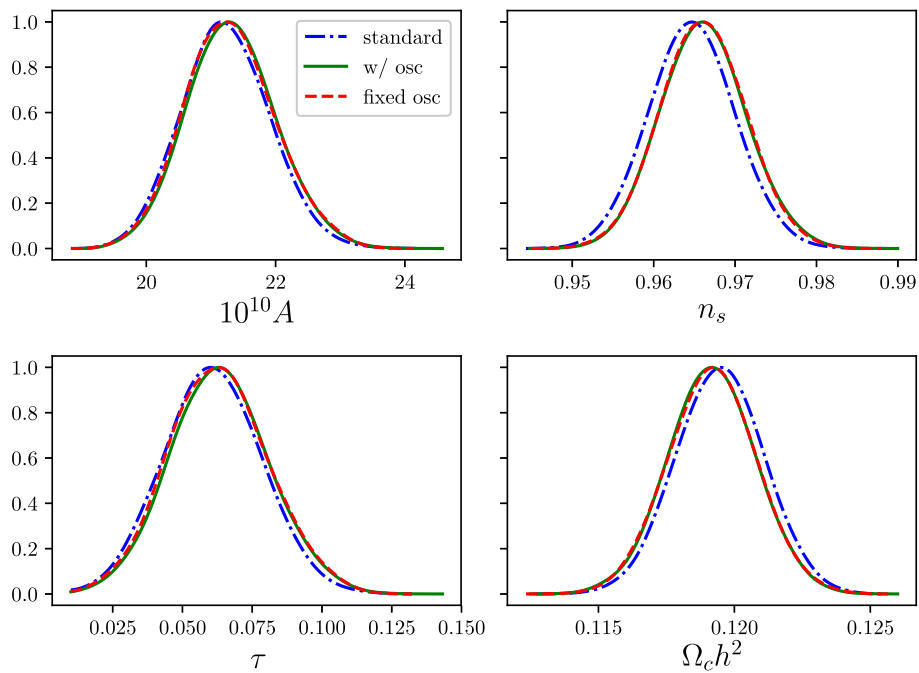}
%%%call your figure name in the place "figurename.eps"
\caption{The posterior distributions of the cosmological parameters $10^9A$ (top left), $n_s$ (top right), $\tau$ (bottom left) and $\Omega_ch^2$ (bottom right). from the standard power law model (blue solid dot and dash line), oscillations (green solid line), and the fixed oscillation model (red dashed line) derived from the TT, TE and EE combined data. }
\label{fig:TTTEEEmulti}
\end{figure}

%\TODO
We have seen the existence of the features in the Planck data. We need
to take into account these features when we estimate the other
cosmological parameters. In fact, the inclusion of the features in the
initial power spectrum has considerably affected the estimation of cosmological parameters if we use the WMAP5 data \cite{2010PhRvD..81h3010I}, especially the estimation of the amplitude of the initial power spectrum $A$ and the spectral index $n_s$. From \tabref{table:cosmo_params_best} and \tabref{table:cosmo_params_mean}, we can check the effects of the features on the estimation of the best fit values and the mean values of cosmological parameters for the Planck data. In these tables, the estimated values, including the amplitude $A$ and the spectral index $n_s$, have barely changed among the models for each data, namely the standard initial power spectrum, the oscillating initial power spectrum and the fixed oscillation initial power spectrum.
From the posterior distributions of some cosmological parameters (\figref{fig:TTTEEEmulti}), 
we can confirm that the distributions do not vary between the different models of initial power spectrum. 
These indicate that the local features in the Planck data do not affect the estimation of cosmological parameters. 
This result is consistent with the analysis for other scales using a similar feature model and Planck data \cite{2014PhRvD..90b3511A,2016arXiv161110350T}.
The improved angular resolution of Planck enables us to observe the greater number of the multipoles  at small scales. 
These rich small scale data has sufficient statistical power to determine the cosmological parameters.
This is why the local features do not affect the estimation of the cosmological parameters.

%-----------------posterior distribution : cosmological parameters
%-----------------TTTEEE

\section{Conclusion}
\label{sec:conclusion}
In this paper, we have studied the features at $\ell \sim 120$ in the Planck2015 CMB angular power spectra data which were obtained by the WMAP5 data analysis \cite{2010PhRvD..81h3010I}. 

We have performed an MCMC analysis using the TT auto-correlation data
and TT, TE, EE combined data of the Planck2015, 
and confirmed  the existence of  the features in the Planck2015 data. 
We checked the consistency between the temperature fluctuation data 
and the polarization data confirming that the features exist in both
cases  with
 almost the same width and at 
the same position albeit a  slightly smaller amplitude in the latter data.
Then we have investigated effects of these features at $\ell\sim120$ in 
determining cosmological parameters and found that they do not affect 
the estimation of cosmological parameters 
unlike the case of WMAP5 data, because there are enough data 
in the higher multipoles to determine the cosmological parameters.

%We could see the features at least in the TT, TE, and EE correlation of the Planck data and they do not affect on the cosmological parameter estimation. 

We expect they are the real features of the initial power spectrum and will be observed by the future experiments without interrupting the estimation of the cosmological parameters.

\section*{Acknowledgment}
This work is supported in part by a Grant-in-Aid for JSPS Research 
under Grant No.\ 15J05029 (K.H.), JSPS KAKENHI Grant-in-Aid for Scientific 
Research Nos.\ 16H01543 (K.I.), 15H02082(J.Y.), 
and Grant-in-Aid for Scientific Research on Innovative Areas 15H05888(J.Y.).


%\bibliographystyle{ptephy}
%\bibliography{sample}
%
% once the .bbl file has been generated then place the text in your article.
\bibliographystyle{ptephy}
\bibliography{sample}

\begin{thebibliography}{10}

\bibitem{Mukhanov:1981xt}
Viatcheslav~F. Mukhanov and G.~V. Chibisov, JETP Lett., {\bf 33}, 532--535,
  [Pisma Zh. Eksp. Teor. Fiz.33,549(1981)] (1981).

\bibitem{1982PhLB..115..295H}
S.~W. {Hawking}, Physics Letters B, {\bf 115}, 295--297 (September 1982).

\bibitem{1982PhRvL..49.1110G}
A.~H. {Guth} and S.-Y. {Pi}, Physical Review Letters, {\bf 49}, 1110--1113
  (October 1982).

\bibitem{1982PhLB..117..175S}
A.~A. {Starobinsky}, Physics Letters B, {\bf 117}, 175--178 (November 1982).

\bibitem{STAROBINSKY198099}
A.A. Starobinsky, Physics Letters B, {\bf 91}(1), 99 -- 102 (1980).

\bibitem{PhysRevD.23.347}
Alan~H. Guth, Phys. Rev. D, {\bf 23}, 347--356 (Jan 1981).

\bibitem{1981MNRAS.195..467S}
K.~{Sato}, Mon. Not. R. Astron. Soc., {\bf 195}, 467--479 (May 1981).

\bibitem{Sato:2015dga}
Katsuhiko Sato and Jun'ichi Yokoyama, Int. J. Mod. Phys., {\bf D24}(11),
  1530025 (2015).

\bibitem{2009ApJS..180..330K}
E.~{Komatsu}, J.~{Dunkley}, M.~R. {Nolta}, C.~L. {Bennett}, B.~{Gold},
  G.~{Hinshaw}, N.~{Jarosik}, D.~{Larson}, M.~{Limon}, L.~{Page}, D.~N.
  {Spergel}, M.~{Halpern}, R.~S. {Hill}, A.~{Kogut}, S.~S. {Meyer}, G.~S.
  {Tucker}, J.~L. {Weiland}, E.~{Wollack}, and E.~L. {Wright}, apjs, {\bf 180},
  330--376 (February 2009),  {{0803.0547}}.

\bibitem{2016A&A...594A..13P}
{Planck Collaboration}, P.~A.~R. {Ade}, N.~{Aghanim}, M.~{Arnaud},
  M.~{Ashdown}, J.~{Aumont}, C.~{Baccigalupi}, A.~J. {Banday}, R.~B.
  {Barreiro}, J.~G. {Bartlett}, and et~al., aap, {\bf 594}, A13 (September
  2016),  {{1502.01589}}.

\bibitem{2010PhRvD..81h3010I}
K.~{Ichiki}, R.~{Nagata}, and J.~{Yokoyama}, Phys.\ Rev.\ D, {\bf 81}(8),
  083010 (April 2010),  {{arXiv:0911.5108}}.

\bibitem{2013JCAP...12..035H}
D.~K. {Hazra}, A.~{Shafieloo}, and G.~F. {Smoot}, JCAP, {\bf 12}, 035 (December
  2013),  {{arXiv:1310.3038}}.

\bibitem{2016JCAP...08..028R}
A.~{Ravenni}, L.~{Verde}, and A.~J. {Cuesta}, jcap, {\bf 8}, 028 (August 2016),
   {{1605.06637}}.

\bibitem{2016arXiv161203490G}
A.~{Gallego Cadavid}, A.~{Enea Romano}, and S.~{Gariazzo}, ArXiv e-prints
  (December 2016),  {{1612.03490}}.

\bibitem{Matsumiya:2001xj}
Makoto Matsumiya, Misao Sasaki, and Jun'ichi Yokoyama, Phys. Rev., {\bf D65},
  083007 (2002),  {{arXiv:astro-ph/0111549}}.

\bibitem{Matsumiya:2002tx}
Makoto Matsumiya, Misao Sasaki, and Jun'ichi Yokoyama, JCAP, {\bf 0302}, 003
  (2003),  {{arXiv:astro-ph/0210365}}.

\bibitem{Kogo:2003yb}
Noriyuki Kogo, Makoto Matsumiya, Misao Sasaki, and Jun'ichi Yokoyama,
  Astrophys. J., {\bf 607}, 32--39 (2004),  {{arXiv:astro-ph/0309662}}.

\bibitem{Kogo:2005qi}
Noriyuki Kogo, Misao Sasaki, and Jun'ichi Yokoyama, Prog. Theor. Phys., {\bf
  114}, 555--572 (2005),  {{arXiv:astro-ph/0504471}}.

\bibitem{2009PhRvD..79d3010N}
R.~{Nagata} and J.~{Yokoyama}, Phys.\ Rev.\ D, {\bf 79}(4), 043010 (February
  2009),  {{0812.4585}}.

\bibitem{2006MNRAS.367.1095T}
D.~{Tocchini-Valentini}, Y.~{Hoffman}, and J.~{Silk}, mnras, {\bf 367},
  1095--1102 (April 2006),  {{astro-ph/0509478}}.

\bibitem{2014A&A...571A..22P}
{Planck Collaboration}, P.~A.~R. {Ade}, N.~{Aghanim}, C.~{Armitage-Caplan},
  M.~{Arnaud}, M.~{Ashdown}, F.~{Atrio-Barandela}, J.~{Aumont},
  C.~{Baccigalupi}, A.~J. {Banday}, and et~al., aap, {\bf 571}, A22 (November
  2014),  {{1303.5082}}.

\bibitem{2013PhRvD..87b3008K}
K.~{Kumazaki}, K.~{Ichiki}, N.~{Sugiyama}, and J.~{Silk}, Phys.\ Rev.\ D, {\bf
  87}(2), 023008 (January 2013),  {{arXiv:1211.3097}}.

\bibitem{2014JCAP...01..025H}
P.~{Hunt} and S.~{Sarkar}, JCAP, {\bf 1}, 025 (January 2014),
  {{arXiv:1308.2317}}.

\bibitem{2014JCAP...11..011H}
D.~K. {Hazra}, A.~{Shafieloo}, and T.~{Souradeep}, JCAP, {\bf 11}, 011
  (November 2014),  {{1406.4827}}.

\bibitem{2011JCAP...01..030A}
A.~{Ach{\'u}carro}, J.-O. {Gong}, S.~{Hardeman}, G.~A. {Palma}, and S.~P.
  {Patil}, JCAP, {\bf 1}, 030 (January 2011),  {{arXiv:1010.3693}}.

\bibitem{2012JCAP...05..008C}
S.~{C{\'e}spedes}, V.~{Atal}, and G.~A. {Palma}, JCAP, {\bf 5}, 008 (May 2012),
   {{arXiv:1201.4848}}.

\bibitem{2012JCAP...10..040G}
X.~{Gao}, D.~{Langlois}, and S.~{Mizuno}, JCAP, {\bf 10}, 040 (October 2012),
  {{arXiv:1205.5275}}.

\bibitem{2012JCAP...11..036S}
R.~{Saito}, M.~{Nakashima}, Y.-i. {Takamizu}, and J.~{Yokoyama}, JCAP, {\bf
  11}, 036 (November 2012),  {{1206.2164}}.

\bibitem{2015JHEP...08..115G}
X.~{Gao} and J.-O. {Gong}, Journal of High Energy Physics, {\bf 8}, 115 (August
  2015),  {{1506.08894}}.

\bibitem{2013JCAP...02..005K}
T.~{Kobayashi} and J.~{Yokoyama}, JCAP, {\bf 2}, 005 (February 2013),
  {{arXiv:1210.4427}}.

\bibitem{2016EPJC...76..385C}
A.~G. {Cadavid}, A.~E. {Romano}, and S.~{Gariazzo}, European Physical Journal
  C, {\bf 76}, 385 (July 2016),  {{1508.05687}}.

\bibitem{2011PThPh.125.1035N}
M.~{Nakashima}, R.~{Saito}, Y.~{Takamizu}, and J.~{Yokoyama}, Progress of
  Theoretical Physics, {\bf 125}, 1035--1052 (May 2011),  {{arXiv:1009.4394}}.

\bibitem{2014PhRvD..90b3511A}
A.~{Ach{\'u}carro}, V.~{Atal}, B.~{Hu}, P.~{Ortiz}, and J.~{Torrado}, prd, {\bf
  90}(2), 023511 (July 2014),  {{1404.7522}}.

\bibitem{2016arXiv161110350T}
J.~{Torrado}, B.~{Hu}, and A.~{Achucarro}, ArXiv e-prints (November 2016),
  {{1611.10350}}.

\bibitem{2016JCAP...09..009H}
D.~K. {Hazra}, A.~{Shafieloo}, G.~F. {Smoot}, and A.~A. {Starobinsky}, JCAP,
  {\bf 9}, 009 (September 2016),  {{1605.02106}}.

\bibitem{2008PhRvD..78l3002N}
R.~{Nagata} and J.~{Yokoyama}, Phys.\ Rev.\ D, {\bf 78}(12), 123002 (December
  2008),  {{0809.4537}}.

\bibitem{2011JCAP...12..008K}
K.~{Kumazaki}, S.~{Yokoyama}, and N.~{Sugiyama}, JCAP, {\bf 12}, 008 (December
  2011),  {{1105.2398}}.

\bibitem{2002PhRvD..66j3511L}
A.~{Lewis} and S.~{Bridle}, Phys.\ Rev.\ D, {\bf 66}(10), 103511 (November
  2002),  {{astro-ph/0205436}}.

\bibitem{akaike1974}
Hirotugu Akaike, Automatic Control, IEEE Transactions on, {\bf 19}(6), 716--723
  (December 1974).

\bibitem{2016JCAP...10..041B}
M.~{Ballardini}, F.~{Finelli}, C.~{Fedeli}, and L.~{Moscardini}, jcap, {\bf
  10}, 041 (October 2016),  {{1606.03747}}.

\end{thebibliography}

%\begin{thebibliography}{9}
%\bibitem{1}
%J. P. Blaizot, and E. Iancu, Phys. Rep. {\bf 359}, 355 (2002).
%\bibitem{2}
%M.~Gyulassy, and L.~McLerran, Nucl.\ Phys.\  A {\bf 750}, 30 (2005).
%\bibitem{3}
%U. W. Heinz, and P. F. Kolb, Nucl. Phys. {\bf A702}, 269 (2002).
%\bibitem{4}
%T.~Hirano, U.~W.~Heinz, D.~Kharzeev, R.~Lacey, and Y.~Nara,
%Phys.\ Lett.\  B {\bf 636}, 299 (2006).
%\bibitem{5}
%R. Baier, A. H. Nueller, D. Schiff, and D. T. Son, Phys. Lett. B {\bf 502}, 51 (2001).
%\end{thebibliography}

%\appendix

%\section{Appendix head}

%This is the sample text. This is the sample text. This is the sample text. This is the sample text.

\end{document}